\documentclass[aps, prd, reprint, superscriptaddress, amsmath, amssymb]{revtex4-2}

\usepackage [pdftex] {graphicx}
\usepackage {epstopdf}

\usepackage{dcolumn}
\usepackage{bm}
\usepackage[utf8]{inputenc}
\usepackage{color,xcolor}

\usepackage{subcaption}

\usepackage{url}
\usepackage{hyperref}
\usepackage{booktabs}
\usepackage{amsmath, graphicx,algorithm,algorithmic,amsfonts,float,multirow}

\usepackage{diagbox}

\begin{document}
	
\preprint{APS/123-QED}

\title{Detection of Multiband Lensed Gravitational Waves from Dark Matter Halos with Deep Learning}

\newcommand{\CQUPhys}{Department of Physics, Chongqing University, Chongqing 401331, P.R. China}
\newcommand{\CQUKeyLab}{Chongqing Key Laboratory for Strongly Coupled Physics, Chongqing University, Chongqing 401331, P.R. China}
\newcommand{\CQUInter}{Institute of Advanced Interdisciplinary Studies, Chongqing University, Chongqing 401331, China}
\newcommand{\XTUInfo}{Department of Electronical Information Science and Technology, Xingtai University, Xingtai 054001, P.R. China}
\newcommand{\SUSTechEarth}{Department of Earth and Sciences, Southern University of Science and Technology, Shenzhen 518055, P.R. China}
\newcommand{\SUSTechPhys}{Department of Physics, Southern University of Science and Technology, Shenzhen 518055, P.R. China}
\newcommand{\CQUMicro}{School of Microelectronics and Communication Engineering, Chongqing University, Chongqing 401331, P.R. China}

\author{Mengfei~Sun}
\affiliation{\CQUPhys}
\affiliation{\CQUKeyLab}

\author{Jie~Wu}
\affiliation{\CQUPhys}
\affiliation{\CQUKeyLab}

\author{Qianning~Hu}
\affiliation{\CQUPhys}
\affiliation{\CQUKeyLab}

\author{Jin~Li}
\email{cqujinli1983@cqu.edu.cn}
\affiliation{\CQUPhys}
\affiliation{\CQUKeyLab}
\affiliation{\CQUInter}

\author{Nan~Yang}
\affiliation{\CQUKeyLab}
\affiliation{\XTUInfo}

\author{Xianghe~Ma}
\affiliation{\CQUPhys}
\affiliation{\CQUKeyLab}

\author{Borui~Wang}
\affiliation{\SUSTechEarth}

\author{Minghui~Zhang}
\affiliation{\SUSTechPhys}

\author{Yuanhong~Zhong}
\email{zhongyh@cqu.edu.cn}
\affiliation{\CQUMicro}

\date{\today}

\begin{abstract}

Lensed gravitational waves acquire amplitude and phase modulations when propagating through the gravitational potential of dark matter halos, producing interference structures in the observed waveform. However, these features are often difficult to identify in detector noise. In this work, we develop a deep learning framework for the automatic classification of lensed gravitational wave signals under multiband observations. We simulate binary neutron star signals observed by the space based detector DECIGO and the ground based Einstein Telescope, and construct five classes of data including pure noise, unlensed signals, and three lensed cases generated by the SIS, CIS, and NFW dark matter halo models. By comparing single detector and joint detector configurations, we evaluate the classification performance under different observational settings. The results show that multiband observations significantly improve the identification of lensed signals and reduce confusion among different lens models. This approach provides an efficient method for automated recognition of lensed gravitational waves in future multiband gravitational wave observations.

\end{abstract}

\maketitle
\section{Introduction}
\label{sec:Introduction}

With the first direct detection of gravitational waves (GWs) from the binary black hole merger GW150914 \citep{abbott2016observation} by the LIGO/Virgo Collaboration, GWs astronomy was established as a powerful probe of the Universe under extreme conditions. Since then, numerous events from binary black holes, binary neutron stars, and black hole–neutron star mergers have been observed~\citep{abbott2017gravitational,ligo2017gravitational,abbott2016gw151226,abbott2017gw170817,abbott2021observation,abbott2020gw190521,abbott2020gw190412}. These detections have verified general relativity in the strong field regime and enabled studies of compact object formation, neutron star equations of state, and cosmological parameters~\citep{schneider2005gravitational,dai2017waveforms,kelly2023constraints,pascale2025sn}. 
Meanwhile, the $\Lambda$ Cold Dark Matter ($\Lambda$CDM) model has become the standard cosmological framework, supported by extensive observations~\citep{alam2021completed,scolnic2018complete,aghanim2020planck,ade2016planck,perlmutter1999measurements,riess1998observational}. Although the $\Lambda$CDM paradigm successfully explains large scale structure formation, several small scale discrepancies remain, such as the anomalous X-ray features in galaxy clusters~\citep{bulbul2014detection} and the paucity of satellite galaxies in the Milky Way~\citep{koposov2008luminosity}. These tensions have motivated alternative dark matter models and efforts to probe dark matter distribution across multiple scales through diverse astronomical observations.  
Within this context, the lensed GWs have emerged as a new frontier. When a gravitational wave passes near a compact object or dark matter substructure, spacetime curvature modulates its waveform, producing amplitude magnification, phase shifts, and interference fringes~\citep{takahashi2003wave,dai2018detecting}. When the wavelength becomes comparable to the lens's Schwarzschild radius, characteristic wave optics effects arise. These imprints provide unique opportunities to detect low-mass dark-matter halos and primordial black holes, offering a novel channel for exploring the nature and distribution of dark matter.

Recent theoretical and numerical studies have shown that lensed GWs introduce distinctive amplitude modulations and time delays in the frequency domain, providing a unique probe of dark matter halos and primordial black holes~\citep{tambalo2023lensing,leung2025wave}. Building on these findings, upcoming third generation detectors such as the Einstein Telescope (ET) and the space based DECi-hertz Interferometer Gravitational wave Observatory (DECIGO) are expected to enable multiband observations that further enhance the detectability and parameter estimation precision of lensed GWs~\citep{abe2025multi,mukherjee2020multimessenger}. 
However, identifying and characterising lensed GWs remain challenging. Traditional matched filtering and Bayesian inference have played a central role in gravitational wave detection and parameter estimation, and continue to provide reliable results across various conditions. As observational sensitivity increases and the volume of detected events grows, these approaches may face practical limitations related to computational efficiency and waveform modeling~\citep{janquart2021fast,li2018gravitational}. In particular, low SNR conditions and interference induced modulations can further complicate signal recovery. These developments highlight the importance of exploring complementary and adaptive analysis frameworks for next generation gravitational wave data.

With the rapid development of deep learning and machine learning techniques, these methods have been progressively introduced into gravitational wave data analysis and lensed GWs identification~\citep{kim2021identification,goyal2021rapid,li2025identification,liu2025identifying,chan2025detectability}. Compared with traditional matched filtering and Bayesian inference approaches, deep learning offers advantages in automatic feature extraction and efficient inference. Convolutional neural networks (CNNs) are capable of learning informative representations from time domain or time frequency domain data, which can help mitigate the dependence on predefined waveform templates. Moreover, their inference phase is computationally efficient, making them particularly suitable for real time identification and filtering of the massive number of GWs events expected in future observations~\citep{singh2018classifying,goyal2021rapid}. Recent studies have demonstrated the potential of deep learning in identifying lensed GWs~\citep{wilde2022detecting}, further confirming its applicability under various conditions.
The advent of third generation detectors has also introduced new opportunities to overcome current limitations through multiband observations. Space based detectors such as the DECIGO exhibits higher sensitivity to low frequency wave optics interference effects, while ground based detectors such as the ET offer superior resolution for geometric optics magnification and phase delay phenomena at mid to high frequencies~\citep{abe2025multi}. This complementarity across frequency bands implies that, by fusing data from multiple detectors, one can simultaneously exploit multiple scale physical features including diffraction and geometric magnification—to mitigate parameter degeneracies and enhance the accuracy of lensed GWs identification. A deep learning based, multiple detector framework for lensed GWs recognition is therefore expected to play a crucial role in future third generation and space based GWs detectors, providing new methodologies for probing dark matter structures and constraining cosmological parameters.

Building on previous studies, this work proposes a deep learning framework for identifying lensed gravitational waves from binary neutron star (BNS) systems under multiband observations. The framework employs a residual convolutional neural network to perform end-to-end classification of time-domain waveforms in detector noise. Two experimental sets are designed. The first constructs single-detector models for the ET and the DECIGO, as well as a joint multiband model (ET+DECIGO), in order to compare classification performance under different observational configurations and evaluate the benefits of multiband observations. The second divides the impact parameter interval $y \in [0,1]$ into five equal-width bins to investigate the dependence of model performance on the lensing geometry. The results show that multiband observations significantly improve the identification of lensed gravitational waves and effectively reduce confusion among different lens models, indicating that multiband data fusion provides an efficient approach for automated recognition of lensed signals in future gravitational-wave observations.

The paper is structured as follows. Section~\ref{sec:Introduction} outlines the scientific motivation and background. Section~\ref{sec:Theoretical Framework} introduces the theoretical framework of lensed gravitational waves and the associated signal modeling. Section~\ref{sec:Detector Configuration and Noise Modeling} describes the detector configurations and noise modeling. Section~\ref{sec:Dataset Construction and Simulation Setup} presents the dataset construction and simulation setup. Section~\ref{sec:Deep Learning Architecture and Training Strategy} details the deep learning architecture and training strategy. Section~\ref{sec:Results and Analysis} reports the classification results under different detector configurations and analyzes the dependence of model performance on the impact parameter $y$. Section~\ref{sec:Conclusion and Outlook} concludes with a summary and future outlook.

\section{Theoretical Framework}
\label{sec:Theoretical Framework}

This section outlines the theoretical framework and modelling approach for describing the propagation of GWs through dark matter halo lenses. It introduces the propagation equation under the weak field approximation and summarises the principal characteristics of the adopted mass distribution models.

\subsection{Fundamental Theory of Lensed GWs}
\label{subsec:Fundamental Theory of Lensed GWs}

In the weak field limit, the propagation of GWs through an inhomogeneous spacetime can be described as a perturbation of the background Friedmann–Lemaître–Robertson–Walker (FLRW) metric. Denoting the background metric as $g^{B}_{\mu\nu}$, the perturbed metric is written as
\begin{align}
g_{\mu\nu}=g^{B}_{\mu\nu}+a^{2}h_{\mu\nu},
\end{align}
where $a$ is the cosmological scale factor and $h_{\mu\nu}$ is the dimensionless tensor perturbation representing the GWs disturbance. The Newtonian potential of the lens, $\Phi(\mathbf{x})$, satisfies $\Phi \ll 1$ to ensure the validity of the linear approximation. Under this condition, the GWs propagation equation can be approximated as a scalar wave equation with a potential term~\citep{takahashi2003wave}:
\begin{align}
\partial_{\mu}\!\left(\!\sqrt{-g^{B}}\,g^{\mu\nu}_{B}\,\partial_{\nu}h\!\right)=0,
\end{align}
where $h(\tau,\mathbf{x})$ is the scalar form of the perturbation. Transforming into the frequency domain yields a Helmholtz-like equation~\citep{peters1974index}:
\begin{align}
\left(\nabla^{2}+\omega^{2}\right)\tilde{h}=4\omega^{2}\Phi\,\tilde{h},
\end{align}
where $\tilde{h}$ is the frequency domain waveform and $\omega=2\pi f$ is the angular frequency.  

To quantify the lensing modulation, the complex amplification factor is defined~\citep{nakamura1999wave,takahashi2003wave}:
\begin{align}
F \equiv \frac{\tilde{h}}{\tilde{h}_{\mathrm{UL}}},
\end{align}
where $\tilde{h}_{\mathrm{UL}}$ denotes the unlensed waveform. Under the thin lens and Born approximations, $F$ can be expressed as a two dimensional integral over the lens plane~\citep{nakamura1999wave,takahashi2003wave}:
\begin{align}
F(w,y)=\frac{w}{2\pi i}\!\int\!d^{2}x\,\exp\!\left[iw\,T(\mathbf{x},\mathbf{y})\right],
\label{eq:amplification_factor}
\end{align}
where $w$ is the dimensionless frequency parameter proportional to both the observing frequency and the lens mass, $y$ is the dimensionless impact parameter, and $T(\mathbf{x},\mathbf{y})$ is the dimensionless time delay function. 
Here $\boldsymbol{\xi}$ and $\boldsymbol{\eta}$ denote the two dimensional position vectors on the lens plane and the source plane, respectively. 
The dimensionless variables are defined as
\begin{align}
x = \frac{\xi}{\xi_{0}}, \qquad 
y = \frac{D_{L}}{\xi_{0}D_{S}}\eta, \qquad
w = \frac{\xi_{0}^{2}}{D_{\mathrm{eff}}}\omega,
\end{align}
where $\xi_{0}$ denotes the characteristic length scale of the lens, and $\omega = 2\pi f$ is the angular frequency of the gravitational wave. 
The effective distance is given by
\begin{align}
D_{\mathrm{eff}} = \frac{D_{L}D_{LS}}{(1+z_{L})D_{S}},
\label{eq:Deff}
\end{align}
where $D_{L}$, $D_{S}$, and $D_{LS}$ are the angular diameter distances from the observer to the lens, from the observer to the source, and from the lens to the source, respectively. 
The quantity $z_{L}$ denotes the redshift of the lens. 
To characterize the lens scale, we define the effective redshifted lens mass as $M_{Lz} = (1+z_{L})M_{L}$, 
and adopt the normalization $\xi_{0} = \sqrt{4 G D_{\mathrm{eff}} M_{Lz}}$, 
so that,
\begin{align}
w = \frac{8\pi G}{c^{3}}\,M_{Lz}\,f.
\label{eq:w}
\end{align}
Hence, $w$ reflects the ratio between the Schwarzschild radius of the lens and the wavelength of the gravitational wave. 
A large $w$ corresponds to the geometrical optics regime ($w \gg 1$), while a small $w$ corresponds to the wave optics regime ($w \ll 1$)~\citep{abe2025multi}. 
In this context, DECIGO mainly operates in the wave optics regime due to its lower frequency, whereas ET approaches the geometrical optics regime at higher frequencies.

The time delay function combines the geometric path difference and the gravitational potential:
\begin{align}
T(\mathbf{x},\mathbf{y})=\tfrac{1}{2}|\mathbf{x}-\mathbf{y}|^{2}-\psi(\mathbf{x})-\phi_{m}(\mathbf{y}),
\end{align}
where $\psi(\mathbf{x})$ is the projected lensing potential, and the constant $\phi_{m}(\mathbf{y})$ normalises the minimum of $T$ to zero. The potential $\psi(\mathbf{x})$ is linked to the surface mass density $\Sigma(\mathbf{x})$ through the two dimensional Poisson equation:
\begin{align}
\nabla^{2}_{\mathbf{x}}\psi(\mathbf{x}) = 2\,\frac{\Sigma(\mathbf{x})}{\Sigma_{\mathrm{cr}}},
\end{align}
where $\Sigma_{\mathrm{cr}}=[4\pi G(1+z_{L})D_{\mathrm{eff}}]^{-1}$ is the critical surface mass density, and $D_{\mathrm{eff}}$ depends on the source lens observer geometry.
The theoretical framework above provide the foundation for the subsequent numerical simulations and data analysis.

\subsection{Lens Models}
\label{subsec:Lens Models}

To investigate the lensed amplification effects of dark matter halos on GWs, three spherically symmetric mass density models are considered: the Singular Isothermal Sphere, the Cored Isothermal Spher, and the Navarro–Frenk–White profile. These models represent a gradual transition from dense to diffuse halo structures, spanning mass distributions from galactic to cluster scales. To focus on the dominant physical effects and maintain analytical tractability, all lenses are assumed to be perfectly spherical, neglecting secondary perturbations and substructure effects~\citep{poon2025galaxy}. The analysis is performed under the thin-lens and weak field approximations, allowing the amplification factor $F(f,y)$ to be derived in a unified formalism.

\paragraph*{\textcolor{black!60!black}{Singular Isothermal Sphere (SIS)}}
The SIS model describes self-gravitating systems with constant velocity dispersion. Its three dimensional density is
\begin{align}
\rho(r)=\frac{v^{2}}{2\pi G r^{2}},
\end{align}
where $v$ is the velocity dispersion. The corresponding surface density is
\begin{align}
\Sigma(x)=\frac{v^{2}}{2G\xi_{0}x}.
\end{align}
When $v^{2}=G\Sigma_{\mathrm{cr}}\xi_{0}$, the scale $\xi_{0}$ equals the Einstein radius, giving $\kappa(x)=1/(2x)$ and lensing potential
\begin{align}
\psi(x)=x.
\end{align}
The frequency domain amplification factor is expressed analytically as a hypergeometric series~\citep{matsunaga2006finite}:
\begin{align}
F(w,y)
&= e^{i w y^{2} / 2}
   \sum_{n=0}^{\infty}
   \frac{\Gamma(1 + n/2)}{n!}
   \left(2 w e^{i 3\pi / 2}\right)^{n/2} \notag \\[4pt]
&\quad \times
   {}_1F_{1}\!\left(
   1+\frac{n}{2};\,1;\,
   -\frac{i w y^{2}}{2}
   \right),
\end{align}
where $w$ is the dimensionless frequency parameter. As $y$ decreases, interference fringes become sharper and magnification stronger.

\paragraph*{\textcolor{black!60!black}{Cored Isothermal Sphere (CIS)}}
The CIS model extends the SIS profile by introducing a finite core radius $r_c$, representing halos with flattened central densities~\citep{kormann1994isothermal,flores1996cluster,treu2010strong}. Its three dimensional and surface densities are
\begin{align}
\rho(r)=\rho_0\,\frac{r_c^2}{r^2+r_c^2},\qquad
\Sigma(x)=\frac{\pi\rho_0 x_c\xi_0}{\sqrt{(x/x_c)^2+1}},
\end{align}
where $x_c=r_c/\xi_0$. For $\rho_0=\Sigma_{\mathrm{cr}}/(2\pi\xi_0x_c^2)$, $\xi_0$ corresponds to the Einstein radius. The lensing potential is
\begin{align}
\psi(x)=\sqrt{x^2+x_c^2}+x_c\ln\!\left(\frac{2x_c}{\sqrt{x^2+x_c^2}+x_c}\right).
\end{align}
Compared with the SIS model, the CIS profile yields a smoother central potential. Increasing $x_c$ weakens focusing and reduces fringe contrast. The amplification factor must be obtained numerically, depending on $(w,y,x_c)$.

\paragraph*{\textcolor{black!60!black}{Navarro–Frenk–White (NFW)}}
The NFW profile, derived from cosmological $N$ body simulations, describes cold dark matter halos~\citep{navarro1996structure,navarro1997universal}:
\begin{align}
\rho(r)=\frac{\rho_s}{(r/r_s)(1+r/r_s)^2}.
\end{align}
Defining $x_s=r_s/\xi_0$, the surface density is
\begin{align}
\Sigma(x)=2\rho_s r_s\,\frac{1-g(x/x_s)}{(x/x_s)^2-1},
\end{align}
where
\begin{align}
g(a)=
\begin{cases}
\dfrac{\arctan\sqrt{a^2-1}}{\sqrt{a^2-1}}, & a>1,\\[6pt]
\dfrac{\mathrm{arctanh}\sqrt{1-a^2}}{\sqrt{1-a^2}}, & a<1,\\
1, & a=1.
\end{cases}
\end{align}
For $\rho_s=\Sigma_{\mathrm{cr}}/(4x_s^3\xi_0)$, the lensing potential becomes
\begin{align}
\psi(x)=\tfrac{1}{2}\!\left[\ln^2\!\left(\frac{x}{2x_s}\right)+\big((x/x_s)^2-1\big)g^2(x/x_s)\right].
\end{align}
The frequency domain properties of the NFW model are sensitive to $r_s$: smaller $r_s$ yields more concentrated halos, stronger magnification, denser fringes, and higher modulation contrast.

In this work, the amplification factors $F(w,y)$ for the three dark matter halo lens models (SIS, CIS, and NFW) were computed using the \texttt{GLoW\_public} library~\citep{GLoW_public}.

\section{Detector Configuration and Noise Modeling}
\label{sec:Detector Configuration and Noise Modeling}

This section outlines the detector configurations and noise models adopted in this study. It first introduces the detector systems and their sensitivity characteristics, followed by discussions of the antenna response functions (Sec.~\ref{subsec:Detector Response Functions}) and the modelling of detector noise (Sec.~\ref{sec:Noise Modeling and Detector Noise Generation}). A unified specification of these components ensures geometric and noise consistency between the DECIGO and the ET, establishing a consistent basis for waveform simulation and signal analysis of lensed GWs.

\subsection{Detector Systems and Sensitivity Characteristics}
\label{subsec:Detector Systems and Sensitivity Characteristics}

This study considers two next generation GWs detectors: the space based DECIGO~\citep{yagi2011detector,kawamura2021current} and the ground based ET~\citep{punturo2010einstein,hild2009xylophone}. These instruments are complementary in sensitivity and operational frequency range, covering approximately $0.1$–$10~\mathrm{Hz}$ for DECIGO and $10$–$1000~\mathrm{Hz}$ for ET.  
The inspiral phase of BNS merger spans frequencies from $0.1$ to $10^3~\mathrm{Hz}$, tracing the system’s evolution from early orbital contraction to the pre-merger stage. DECIGO is thus sensitive to the low frequency early inspiral, whereas ET captures the mid to high frequency band preceding coalescence. The noise power spectral density (PSD) for DECIGO follows the official design specification~\citep{yagi2011detector}, and that for ET is based on the ET sensitivity curve~\citep{hild2011sensitivity}.  

\begin{figure}
    \centering
    \begin{subfigure}[b]{0.45\textwidth}
        \centering
        \includegraphics[width=\textwidth]{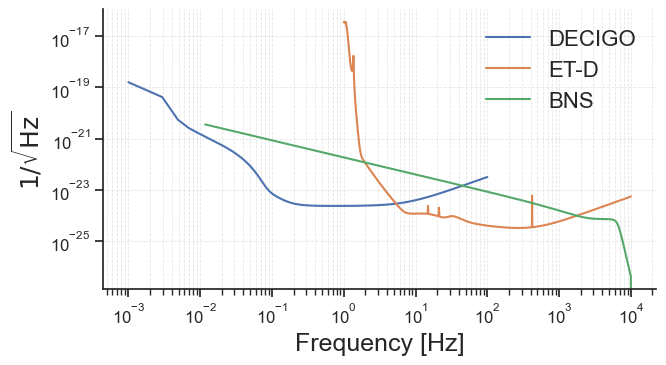}
\caption{Amplitude Spectral Density (ASD) of DECIGO and ET compared with the BNS signal with $m_1=m_2=1.4\,M_\odot$ at 1000\,Mpc ($z=0.203$).}
        \label{subfig:sensitivity_curve_decigo_et_bns}
    \end{subfigure}
    \hfill
    \begin{subfigure}[b]{0.4\textwidth}
        \centering
        \includegraphics[width=\textwidth]{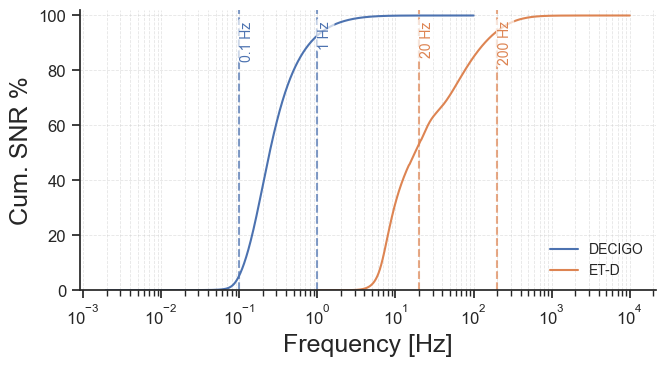}
        \caption{Upper panel waveform used to compute cumulative SNR \% in DECIGO and ET. Total SNRs are $\mathrm{SNR}=238.38$ (DECIGO) and $\mathrm{SNR}=50.70$ (ET). Dashed lines mark the dataset frequency bands (0.1--10 Hz for DECIGO and 20--200 Hz for ET).}
        \label{subfig:cumulative_snr2_fraction_decigo_et}
    \end{subfigure}
    \caption{(a) ASD of DECIGO and ET with BNS characteristic strain; (b) cumulative SNR for the same waveform in both detectors.}
    \label{fig:decigo_et_sensitivity_and_cumulative_snr}
\end{figure}

Fig.~\ref{fig:decigo_et_sensitivity_and_cumulative_snr} compares the sensitivity of DECIGO and ET with the amplitude spectral density (ASD) of a BNS signal. Panel (a) shows the detector ASD curves together with the BNS characteristic strain, while panel (b) presents the cumulative SNR contributions within the adopted frequency bands for the same waveform. DECIGO provides higher sensitivity in the $0.1$–$10~\mathrm{Hz}$ band, where wave–optics effects from low-mass lenses produce interference and diffraction features. In contrast, ET is most sensitive in the $20$–$200~\mathrm{Hz}$ range, where geometric–optics amplification and phase shifts dominate. The complementary frequency coverage enables the same merger event to be observed at different inspiral stages, providing a more complete waveform and improving the identification of lensing signatures.

\subsection{Detector Response Functions}
\label{subsec:Detector Response Functions}

In multiband observations, the detector response functions describe how the detected strain couples to the source direction, polarization, and detector orientation. For an arbitrary GW source, the measured strain at the detector is expressed as
\begin{align}
h_{\mathrm{det}}(t)=F_{+}(t)\,h_{+}(t)+F_{\times}(t)\,h_{\times}(t),
\end{align}
where $F_{+}$ and $F_{\times}$ are the antenna pattern functions for the “$+$” and “$\times$” polarization modes, respectively, and $h_{+}$ and $h_{\times}$ denote the corresponding waveform components. These response functions encode the time dependent coupling between the incident direction $(\theta,\phi)$, the polarization angle $\psi$, and the detector geometry.

DECIGO is a space based interferometer with a triangular three arm configuration whose antenna response varies periodically with time. Let $(\bar{\theta}_{S},\bar{\phi}_{S})$ denote the source position in the heliocentric coordinate system and $(\hat{\theta}_{L},\hat{\phi}_{L})$ the orientation of the orbital angular momentum. The observed strain in the detector frame is~\citep{yagi2011detector}
\begin{align}
h(t)=F_{+}(\bar{\theta}_{S},\bar{\phi}_{S},\hat{\theta}_{L},\hat{\phi}_{L})\,h_{+}(t)
     +F_{\times}(\bar{\theta}_{S},\bar{\phi}_{S},\hat{\theta}_{L},\hat{\phi}_{L})\,h_{\times}(t),
\end{align}
where
\begin{align}
F_{+}(\theta_{S},\phi_{S},\psi_{S})
&=\frac{1}{2}\!\left(1+\cos^{2}\theta_{S}\right)
   \cos(2\phi_{S})\cos(2\psi_{S}) \notag \\[3pt]
&\quad -\,\cos\theta_{S}\sin(2\phi_{S})\sin(2\psi_{S}), \\[5pt]
F_{\times}(\theta_{S},\phi_{S},\psi_{S})
&=\frac{1}{2}\!\left(1+\cos^{2}\theta_{S}\right)
   \cos(2\phi_{S})\sin(2\psi_{S}) \notag \\[3pt]
&\quad +\,\cos\theta_{S}\sin(2\phi_{S})\cos(2\psi_{S}).
\end{align}

The transformation between the angular parameters $(\theta_{S},\phi_{S},\psi_{S})$ and the heliocentric coordinates is given by
\begin{align}
\theta_{S}(t)
 &=\cos^{-1}\!\left[\frac{1}{2}\cos\bar{\theta}_{S}
   -\frac{\sqrt{3}}{2}\sin\bar{\theta}_{S}\cos\!\bigl(\hat{\phi}(t)-\bar{\phi}_{S}\bigr)\right], \\[3pt]
\phi_{S}(t)
 &=\frac{\pi}{12}
   +\tan^{-1}\!\left[
   \frac{\sqrt{3}\cos\bar{\theta}_{S}
        +\sin\bar{\theta}_{S}\cos\!\bigl(\hat{\phi}(t)-\bar{\phi}_{S}\bigr)}
        {2\sin\bar{\theta}_{S}\sin\!\bigl(\hat{\phi}(t)-\bar{\phi}_{S}\bigr)}\right], \\[3pt]
\psi_{S}&=\tan^{-1}\!\left(\frac{a}{b}\right),
\end{align}
where
\begin{align}
a
&= \frac{1}{2}\cos\hat{\theta}_{L}
   -\frac{\sqrt{3}}{2}\sin\hat{\theta}_{L}
      \cos[\hat{\phi}(t)-\hat{\phi}_{L}] \notag \\[3pt]
&\quad
   -\cos\hat{\theta}_{L}\cos^{2}\bar{\theta}_{S}
   -\sin\hat{\theta}_{L}\sin\bar{\theta}_{S}
      \cos(\hat{\phi}_{L}-\bar{\phi}_{S}),
   \label{eq:decigo_angle_a}\\[5pt]
b
&= \frac{1}{2}\sin\hat{\theta}_{L}\sin\bar{\theta}_{S}
      \sin(\hat{\phi}_{L}-\bar{\phi}_{S}) \notag \\[3pt]
&\quad
   -\frac{\sqrt{3}}{2}\cos\hat{\phi}(t)
      [\cos\hat{\theta}_{L}\sin\bar{\theta}_{S}\sin\bar{\phi}_{S}
      -\cos\bar{\theta}_{S}\sin\hat{\theta}_{L}\sin\hat{\phi}_{L}] \notag \\[3pt]
&\quad
   -\frac{\sqrt{3}}{2}\sin\hat{\phi}(t)
      [\cos\bar{\theta}_{S}\sin\hat{\theta}_{L}\cos\hat{\phi}_{L}
      -\cos\hat{\theta}_{L}\sin\bar{\theta}_{S}\cos\bar{\phi}_{S}],
   \label{eq:decigo_angle_b}
\end{align}
and
\begin{align}
\bar{\theta}(t)=\pi/2, \qquad
\hat{\phi}(t)=2\pi t/T+c_{0},
\end{align}
where $T = 1~\mathrm{yr}$ is the orbital period and $c_{0} = 0$ is the initial orbital phase.

The Einstein Telescope consists of three co-located interferometers arranged in an equilateral triangle, each having equal arm lengths and an opening angle of $\gamma = \pi/3$. 
Let $(\theta,\phi)$ denote the source sky location in the detector frame and $\psi$ the polarization angle. 
For a single interferometer with arm orientation offset $\Delta$, the antenna pattern functions for the plus and cross polarizations are given by \citep{hild2011sensitivity}
\begin{align}
F_{+}(\theta,\phi,\psi;\gamma,\Delta)
&= -\frac{\sin\gamma}{2}\!
   \Bigl[
   (1+\cos^{2}\theta)\sin(2\phi')\cos(2\psi) \notag \\[3pt]
&\quad
   +\,2\cos\theta\cos(2\phi')\sin(2\psi)
   \Bigr], \\[5pt]
F_{\times}(\theta,\phi,\psi;\gamma,\Delta)
&= +\frac{\sin\gamma}{2}\!
   \Bigl[
   (1+\cos^{2}\theta)\sin(2\phi')\sin(2\psi) \notag \\[3pt]
&\quad
   -\,2\cos\theta\cos(2\phi')\cos(2\psi)
   \Bigr],
\end{align}
where $\phi' = \phi + \Delta$. 
In the ET-D configuration, the three interferometers are oriented with relative rotation angles 
$\Delta = 0$, $+2\pi/3$, and $-2\pi/3$, 
corresponding to three $60^{\circ}$ separated arms forming an equilateral triangle layout underground. 
Each interferometer therefore provides an independent response following the same functional form but rotated in azimuth by $\Delta$.

For data generation, the directional response is averaged over random sky positions and polarization angles, approximating an isotropic source distribution. To ensure geometric consistency between DECIGO and ET for the same GW source, a reference epoch $t_{\mathrm{ref}}$ in the heliocentric coordinate system is defined, and the polarization angle $\psi$ is fixed accordingly using the DECIGO transformation (equations~\eqref{eq:decigo_angle_a}--\eqref{eq:decigo_angle_b}) with $t=t_{\mathrm{ref}}$:
\begin{align}
\label{eq:jihuajiao}
\bar{\phi}_{t} = \frac{2\pi}{T}\,t_{\rm ref} + c_{0},\qquad
\psi = \tan^{-1}\!\left(\frac{a}{b}\right),
\end{align} 
In the ET response, the parameters are then set to
\begin{align}
\theta=\bar{\theta}_{S}, \qquad
\phi=\bar{\phi}_{S}, \qquad
\psi=\psi(t_{\rm ref}),
\end{align}
where $(\bar{\theta}_{S},\bar{\phi}_{S})$ are the source sky coordinates, and $\psi(t_{\mathrm{ref}})$ is the polarization angle determined by the DECIGO geometry at the reference epoch. 
Here the reference time is set to $t_{\mathrm{ref}}=0$, corresponding to the initial orbital phase of DECIGO. 
This configuration guarantees that both detectors share identical source direction and polarization orientation when observing the same GW event, ensuring full geometric consistency.

\subsection{Noise Modeling and Detector Noise Generation}
\label{sec:Noise Modeling and Detector Noise Generation}

Based on the one sided noise PSD, we could obtain the time domain noise signal from the one sided PSD; in this paper, we utilized the Python function \texttt{pycbc.noise.gaussian.noise\_from\_psd} \citep{alex_nitz_2024_10473621}, which takes a PSD as input and returns colored Gaussian noise, to simulate the time domain noise received by both the DECIGO and ET detectors. The corresponding PSDs of ET and DECIGO were adopted according to their official sensitivity curves \citep{yagi2011detector,hild2011sensitivity}, ensuring that the generated noise realizations accurately reflect the instrumental characteristics across their respective frequency bands.

\section{Dataset Construction and Simulation Setup}
\label{sec:Dataset Construction and Simulation Setup}

Under the thin lens and weak field approximations, the lensed GWs can be modelled as a frequency domain modulation of the unlensed (UL) signal. The UL polarization components $\tilde{h}_{+}^{\rm UL}(f)$ and $\tilde{h}_{\times}^{\rm UL}(f)$ are first generated, and the corresponding lensed signals are obtained by multiplying them by the amplification factor $F(w,y)$:
\begin{align}
\tilde{h}_{+}(f) = F(w,y)\,\tilde{h}_{+}^{\rm UL}(f), \qquad
\tilde{h}_{\times}(f) = F(w,y)\,\tilde{h}_{\times}^{\rm UL}(f),
\end{align}
where $\tilde{h}_{+}(f)$ and $\tilde{h}_{\times}(f)$ denote the “$+$” and “$\times$” polarization components after lensing modulation, and $\tilde{h}_{+}^{\rm UL}(f)$ and $\tilde{h}_{\times}^{\rm UL}(f)$ are their unlensed counterparts. The amplification factor $F(w,y)$ depends on the adopted lens model and includes three classes in this study:
\begin{align}
F(w,y) \in \{F_{\rm SIS}(w,y),\,F_{\rm CIS}(w,y),\,F_{\rm NFW}(w,y)\},
\end{align}
corresponding to the SIS, CIS, and NFW dark matter halo models.

Applying an inverse Fourier transform to the modulated frequency domain signals yields the time domain polarization waveforms $h_{+}(t)$ and $h_{\times}(t)$. Considering the detector antenna responses, the measured strain is given by
\begin{align}
h(t) = F_{+}(t;\Omega,\psi)\,h_{+}(t) + F_{\times}(t;\Omega,\psi)\,h_{\times}(t),
\end{align}
where $F_{+}$ and $F_{\times}$ are the response functions to the two polarizations, $\Omega$ denotes the sky location of the source, and $\psi$ is the polarization angle.  
Finally, detector noise $n(t)$ consistent with the detector PSD $S_n(f)$ is added to obtain the simulated observation:
\begin{align}
s(t) = h(t) + n(t),
\end{align}
where $s(t)$ is the final noisy observation and $n(t)$ is a Gaussian noise sequence with spectral structure defined by $S_n(f)$.  
This procedure comprises four stages, including lensing modulation, Fourier transformation, detector response mapping, and noise injection.

\begin{figure}
\centering
\includegraphics[width=0.35\textwidth]{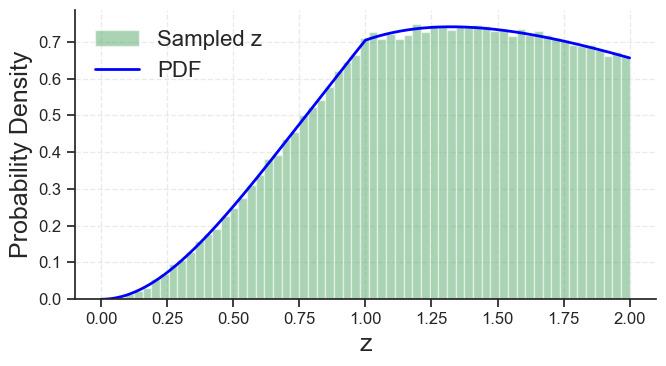}
\caption{BNS z distribution.}
\label{fig:z_distribution}
\end{figure}

The redshift distribution of BNS in our dataset follows the probability density function \cite{18,Sun:2023eic}:
\begin{equation}
\begin{aligned}
\rho(z) \sim \frac{4 \pi d_{C}^{2}(z) R(z)}{H(z)(1+z)},
\end{aligned}
\label{eq:redshift_pdf}
\end{equation}
where the co-moving distance is 
\begin{equation}
d_{C}(z) = \int_{0}^{z} \frac{1}{H\left(z^{\prime}\right)} d z^{\prime}.
\label{eq:comoving_distance}
\end{equation}
The luminosity distance $d_L$ is defined as $d_L = d_C (1+z)$. Other related parameter settings follow \cite{Sun:2023eic}.
The source evolution rate is given by \cite{19,20,21}:
\begin{equation}
\begin{aligned}
R(z)=
\left\{
\begin{array}{cc}
1+2 z, & z \leq 1 \\
\frac{3}{4}(5-z), & 1<z<5. \\
0, & z \geq 5
\end{array}
\right.
\end{aligned}
\label{eq:source_evolution}
\end{equation}
The resulting BNS redshift distribution in our dataset is shown in Fig.~\ref{fig:z_distribution}.

\begin{table}
\centering
\caption{
Sampling ranges of the dataset. 
The angular parameters $(\bar{\theta}_{S},\bar{\phi}_{S},t_{\mathrm{ref}})$ 
correspond to the DECIGO response geometry, where the polarization angle 
$\psi$ is computed at $t_{\mathrm{ref}}$ according to Eq.~\eqref{eq:jihuajiao}. 
For the ET response, the parameters are determined as 
$(\theta,\phi,\psi)_{\mathrm{ET}}=(\bar{\theta}_{S},\bar{\phi}_{S},\psi)$ 
to ensure geometric consistency.
}
\label{tab:param_range}
\begin{tabular}{ccc}
\hline
Parameter & Symbol & Range \\
\hline
Primary mass & $m_{1}$ & $\mathcal{U}(1,3)~M_{\odot}$ \\
Secondary mass & $m_{2}$ & $\mathcal{U}(1,3)~M_{\odot}$ \\
Luminosity distance & $d_{L}$ & from $z$ (Eq.~\eqref{eq:redshift_pdf}) \\
Ecliptic latitude & $\bar{\theta}_{S}$ & $\cos^{-1}\!\bigl(\mathcal{U}(-1,1)\bigr)$ \\
Ecliptic longitude & $\bar{\phi}_{S}$ & $\mathcal{U}(0,2\pi)$ \\
Inclination & $\iota$ & $1.0$ \\
Initial phase & $\phi_{0}$ & $0$ \\
Spin 1 & $\chi_{1}$ & $0.1$ \\
Spin 2 & $\chi_{2}$ & $0.1$ \\
Reference epoch & $t_{\mathrm{ref}}$ & $0$ \\
Impact parameter & $y$ & $\mathcal{U}(0,1)$ \\
Lens mass & $M_{L}$ & $2\times10^{3}~M_{\odot}$ \\
Lens redshift & $z_{L}$ & $0.3$ \\
\hline
\end{tabular}
\end{table}

\begin{table}
\centering
\caption{Frequency domain construction and corresponding time domain forms of the five sample classes.}
\label{tab:classes}
\begin{tabular}{ccc}
\hline
Class & frequency domain  & time domain  \\
\hline
PN & - & $n(t)$ \\
UL & $\tilde{h}^{\rm UL}(f)$ & $h(t) + n(t)$ \\
SIS lens & $F_{\rm SIS}(w,y)\,\tilde{h}^{\rm UL}(f)$ & $h^{\rm SIS}(t) + n(t)$ \\
CIS lens & $F_{\rm CIS}(w,y)\,\tilde{h}^{\rm UL}(f)$ & $h^{\rm CIS}(t) + n(t)$ \\
NFW lens & $F_{\rm NFW}(w,y)\,\tilde{h}^{\rm UL}(f)$ & $h^{\rm NFW}(t) + n(t)$ \\
\hline
\end{tabular}
\end{table}

For both the space based DECIGO detector and the ground based ET detector, five classes of samples were generated: pure noise (PN), unlensed (UL), and three lensed cases (SIS, CIS, NFW). Each class contains $2\times10^5$ samples in each detector dataset, resulting in a total of $10^6$ samples per detector. All physical parameters of the binary neutron star (BNS) systems and lens configurations are summarized in Table~\ref{tab:param_range}, and the definitions of all sample classes are listed in Table~\ref{tab:classes}. DECIGO operates in the $0.1$–$10~\mathrm{Hz}$ frequency band with a sampling rate of $1~\mathrm{Hz}$ and a signal duration of $3600~\mathrm{s}$, while ET covers the $20$–$200~\mathrm{Hz}$ band with a sampling rate of $128~\mathrm{Hz}$ and a duration of $10~\mathrm{s}$. The waveform templates were generated based on the post-Newtonian model described in Refs.~\cite{wu2024comparison,blanchet2024post}, adopting a $2.5$PN approximation including spin–orbit (SO) coupling while neglecting spin–spin effects. The frequency-domain amplification factor $F(w,y)$ was computed numerically using the \texttt{GLoW\_public} package~\citep{GLoW_public}. The lens redshift was fixed at $z_L=0.3$, consistent with the median redshift of spectroscopically confirmed lensing galaxy clusters reported in Ref.~\cite{smith2018if}. In addition, we fixed the lens mass to $M_L = 2\times10^3\,M_\odot$, since this choice allows the wave-optics regime in DECIGO and the geometric-optics regime in ET to be analyzed in the multiband configuration.

\begin{figure}
    \centering
    \begin{subfigure}[b]{0.36\textwidth}
        \centering
        \includegraphics[width=\textwidth]{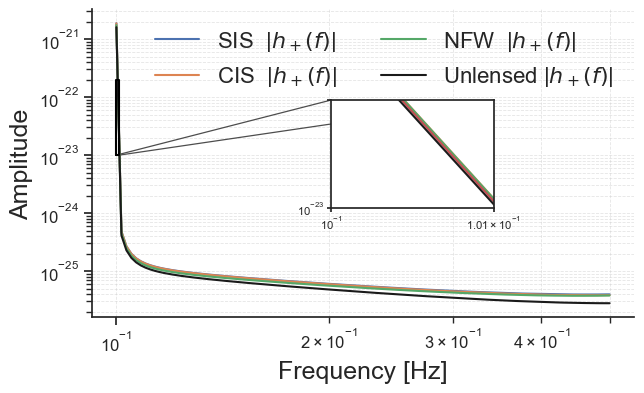}
        \caption{Frequency domain waveforms of DECIGO.}
        \label{subfig:freq_decigo}
    \end{subfigure}
    \hfill
    \begin{subfigure}[b]{0.36\textwidth}
        \centering
        \includegraphics[width=\textwidth]{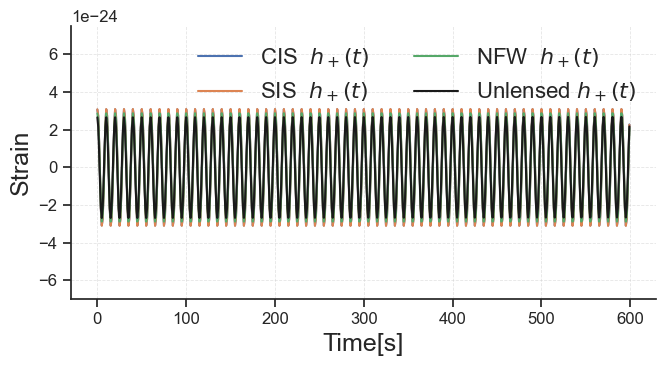}
        \caption{Time domain waveforms of DECIGO.}
        \label{subfig:time_decigo}
    \end{subfigure}
    \caption{Frequency and time domain waveforms of BNS ($m_{1}=m_{2}=1.4\,M_\odot$, $d_{L}=100~\mathrm{Mpc}$)
 under SIS, CIS, and NFW lensing models, as observed by DECIGO. The lens redshift is $z_{L}=0.3$, and the lens parameters are $M_{L}=2\times10^{3}\,M_\odot$, $\psi_{0}=1.0$, $r_{c}=0.3$, $r_{s}=0.3$, and $y=0.3$.}
    \label{fig:sample_decigo}
\end{figure}

\begin{figure}
    \centering
    \begin{subfigure}[b]{0.36\textwidth}
        \centering
        \includegraphics[width=\textwidth]{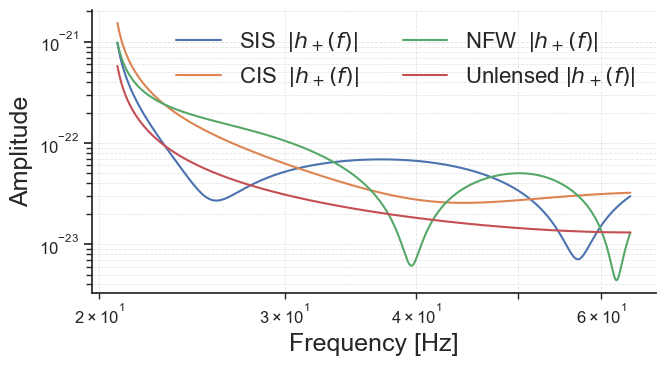}
        \caption{Frequency domain waveforms of ET.}
        \label{subfig:freq_etd}
    \end{subfigure}
    \hfill
    \begin{subfigure}[b]{0.36\textwidth}
        \centering
        \includegraphics[width=\textwidth]{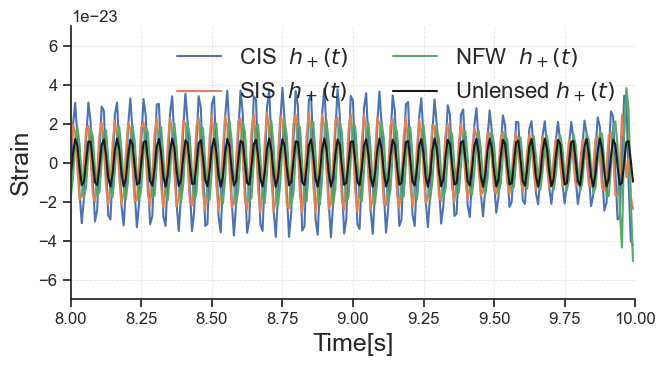}
        \caption{Time domain waveforms of ET.}
        \label{subfig:time_etd}
    \end{subfigure}
    \caption{Frequency and time domain waveforms of ($m_{1}=m_{2}=1.4\,M_\odot$, $d_{L}=100~\mathrm{Mpc}$) under SIS, CIS, and NFW lensing models, as observed by ET. The lens redshift is $z_{L}=0.3$, and the lens parameters are $M_{L}=2\times10^{3}\,M_\odot$, $\psi_{0}=1.0$, $r_{c}=0.3$, $r_{s}=0.3$, and $y=0.3$.}
    \label{fig:sample_etd}
\end{figure}

To provide a clearer illustration of the data structure and model inputs, examples of different sample classes in both the frequency and time domains are shown in Fig.~\ref{fig:sample_decigo} and Fig.~\ref{fig:sample_etd}. These figures display the SIS, CIS, and NFW lensing cases for DECIGO and ET, highlighting the distinct lensing features and their different manifestations across the two frequency bands. Detector noise was generated according to the power spectral density specific to each detector. After applying lensing modulation in the frequency domain, inverse Fourier transformation was performed to obtain the time-domain signals, and independent Gaussian noise realizations were added to each waveform. All signals were adjusted to fixed lengths within each detector dataset to ensure consistent temporal alignment.

\section{Deep Learning Architecture and Training Strategy}
\label{sec:Deep Learning Architecture and Training Strategy}

To enable efficient and robust classification of dark matter halo lensing effects, we use the residual stacked one dimensional convolutional neural network (ResNet-like 1D CNN) that operates directly on time domain waveform inputs. The model classifies five classes: PN, UL signals, and three lensed cases corresponding to the SIS, CIS, and NFW models. By processing standardized time domain strain sequences without frequency domain transformation or handcrafted feature extraction, the network retains temporal continuity and local interference structures, which are expected to contribute to better recognition of weak modulations.

\begin{figure}
\centering
\includegraphics[width=0.35\textwidth]{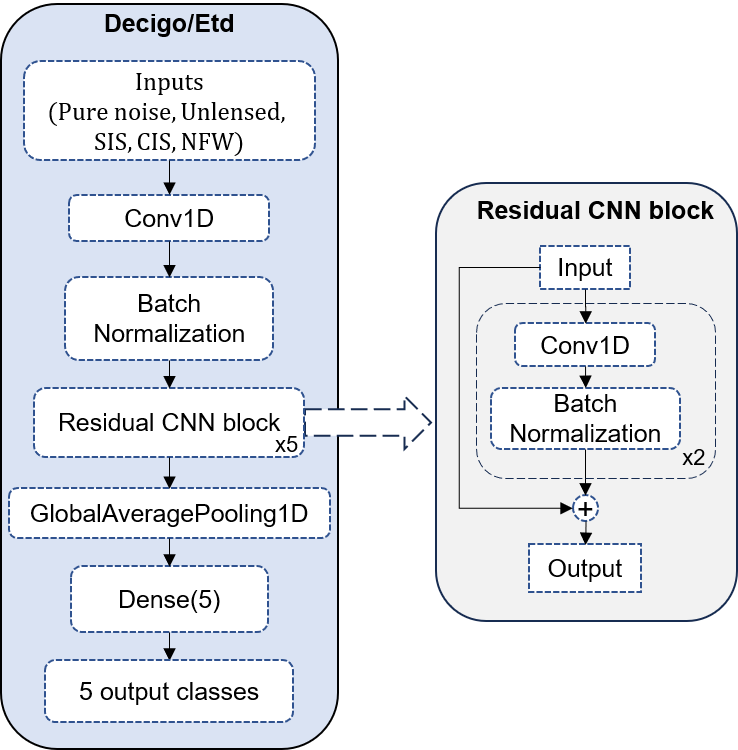}
\caption{Architecture of the residual stacked convolutional network.}
\label{fig:model_single}
\end{figure}

The architecture comprises multiple residual convolutional modules, each containing two convolution–normalization–activation (ReLU) layers with identity shortcuts to facilitate feature propagation and alleviate vanishing gradient or degradation effects. The input first passes through an initial 1D convolutional layer for primary feature extraction, followed by five stacked residual blocks forming deep hierarchical channels. Each block applies a $3\times1$ convolution kernel, Batch Normalization, and a ReLU activation. A global average pooling layer compresses the temporal dimension, and a fully connected layer with a Softmax activation outputs the five class probability distribution. The complete single branch network functions as an end to end multiple class classifier (Fig.~\ref{fig:model_single}).

\begin{figure}
\centering
\includegraphics[width=0.4\textwidth]{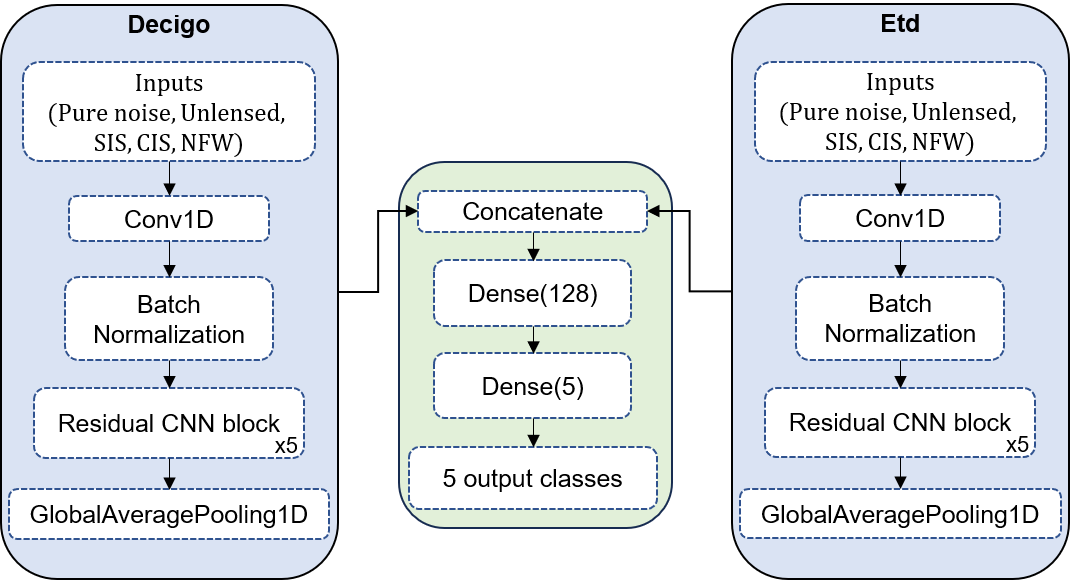}
\caption{A dual branch network that combines features from two detectors (ET and DECIGO) using late fusion.}
\label{fig:model_joint}
\end{figure}

To evaluate the benefit of multiband observations, two network configurations were implemented. The first is a single branch network processing input waveforms from either DECIGO or ET individually. The second is a dual branch fusion network designed for multiple detector inputs, where each branch independently processes data from one detector using identical ResNet substructures. Deep feature maps from both branches are concatenated in feature space and fused via a fully connected layer to form a unified representation. This late fusion approach preserves detector specific feature learning while leveraging the complementarity between the low frequency (DECIGO) and mid to high frequency (ET) bands, thereby enhancing discrimination of both interference–diffraction patterns and geometric magnification effects (Fig.~\ref{fig:model_joint}).

All input waveforms were standardized to zero mean and unit variance to improve numerical stability and convergence. The model was optimized using sparse categorical cross entropy loss and the Adam optimizer with an initial learning rate of $10^{-3}$. The \texttt{ReduceLROnPlateau} strategy was employed to decrease the learning rate adaptively when the validation loss stagnated, and \texttt{EarlyStopping} was used to terminate training when no improvement was observed for several epochs. Training was performed for up to 200 epochs with a batch size of 128 and a 10\% validation split. All experiments used fixed length waveform sequences and output labels for five classes. Model training and inference were accelerated using an NVIDIA A800 80G GPU and two Intel(R) Xeon(R) Silver 8480 CPUs.

\section{Results and Analysis}
\label{sec:Results and Analysis}

In this section, we first present the overall classification performance under single-detector and multiband configurations, and then investigate the dependence of model performance on the impact parameter $y$. To ensure mathematical clarity and reproducibility, model performance is quantified using Accuracy, Precision, Recall, $F_1$ score, and Averaging metrics.

Accuracy measures the overall correctness of classification and is defined as
\begin{align}
\mathrm{Accuracy} = \frac{TP + TN}{TP + TN + FP + FN},
\end{align}
where $TP$, $TN$, $FP$, and $FN$ denote true positives, true negatives, false positives, and false negatives, respectively. Although the above definition corresponds to binary classification, in the present five-class task the reported Accuracy represents the global correct classification rate computed across all classes, characterizing the overall discriminative capability of the model. Precision and Recall quantify the reliability and completeness of predictions and are defined as
\begin{align}
\mathrm{Precision} = \frac{TP}{TP + FP}, \qquad
\mathrm{Recall} = \frac{TP}{TP + FN}.
\end{align}
The $F_1$ score is the harmonic mean of Precision and Recall,
\begin{align}
F_1 = \frac{2\,\mathrm{Precision}\times \mathrm{Recall}}{\mathrm{Precision} + \mathrm{Recall}}.
\end{align}
For multi-class classification, to prevent any class from dominating the evaluation due to sample imbalance, we apply Averaging across all classes, defined as
\begin{align}
\mathrm{Average(M)} = \frac{1}{C}\sum_{i=1}^{C} M_i,
\end{align}
where $C$ is the total number of classes and $M_i$ denotes the metric corresponding to class $i$. This definition ensures equal weight for each lensing class in the overall assessment.

To further analyze inter-class discrimination, confusion matrices are plotted to visualize misclassification relationships among classes. In addition, Receiver Operating Characteristic (ROC) curves are constructed to evaluate the trade-off between the True Positive Rate (TPR) and the False Positive Rate (FPR). The Area Under the Curve (AUC) quantifies the discriminative capability of the model and is defined as
\begin{align}
\mathrm{AUC} = \int_{0}^{1} \mathrm{TPR}(\mathrm{FPR})\, d(\mathrm{FPR}),
\end{align}
with larger AUC values indicating stronger separation ability. Besides ROC curves, we also construct Precision--Recall (PR) curves to characterize the relationship between Precision and Recall. PR curves are particularly informative in classification tasks where the quality of positive predictions is crucial. The area under the PR curve is defined as
\begin{align}
\mathrm{PR\text{-}AUC} = \int_{0}^{1} \mathrm{Precision}(\mathrm{Recall})\, d(\mathrm{Recall}),
\end{align}
where larger values indicate that high precision is maintained while achieving high recall. For the multi-class task, micro averaging is adopted when computing PR-based statistics to provide a global measure of performance.
Regarding the construction of the test data, to systematically investigate the influence of the impact parameter $y$, we divide the interval $y \in [0,1]$ into five equal bins. For each bin, an independent test dataset is newly generated, containing $10^4$ samples per class with $y$ uniformly sampled within the corresponding interval. The five bin datasets are subsequently merged to form the overall test set covering the full range $y \in [0,1]$, which is used to evaluate model performance under complete marginalization over the impact parameter.

\subsection{Overall Classification Performance}
\label{sec:overall_performance}

\begin{figure}
\centering
\includegraphics[width=0.45\textwidth]{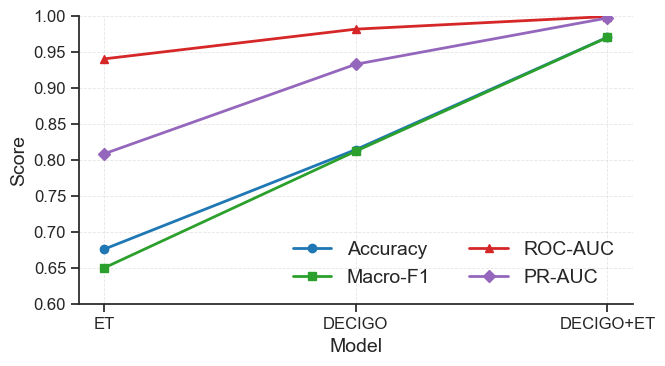}
\caption{
Overall classification performance for ET, DECIGO, and the joint DECIGO+ET models over the full impact parameter range $y \in [0,1]$. The comparison includes Accuracy, Average $F_1$, ROC-AUC, and PR-AUC metrics.
}
\label{fig:there_model_metrics}
\end{figure}

In this subsection, the overall performance is evaluated under complete marginalization over the impact parameter $y \in [0,1]$, following the metrics defined in~\ref{sec:Results and Analysis}, including Accuracy, Precision, Recall, $F_1$, and Macro Average. The metric comparison is shown in~\ref{fig:there_model_metrics}, and the detailed statistics are listed in~\ref{tab:overall_metrics_singlecol}. The test set is constructed by merging the five $y$ bins, covering the full parameter space.

\begin{table}
\centering
\caption{Overall classification performance for single and multiband models with $y\in[0,1]$.}
\label{tab:overall_metrics_singlecol}
\begin{tabular}{lcccc}
\hline
Model & Class & Prec. & Rec. & $F_1$ \\
\hline
\multirow{7}{*}{ET}
 & PN  & 1.000 & 0.990 & 0.995 \\
 & UL  & 0.810 & 0.952 & 0.875 \\
 & SIS & 0.485 & 0.172 & 0.254 \\
 & CIS & 0.441 & 0.554 & 0.491 \\
 & NFW & 0.581 & 0.709 & 0.639 \\
 & Average & 0.663 & 0.675 & 0.651 \\
 & Accuracy  & \multicolumn{3}{c}{0.677} \\
\hline
\multirow{7}{*}{DECIGO}
 & PN  & 1.000 & 1.000 & 1.000 \\
 & UL  & 1.000 & 0.999 & 1.000 \\
 & SIS & 0.531 & 0.597 & 0.562 \\
 & CIS & 0.540 & 0.471 & 0.503 \\
 & NFW & 0.995 & 1.000 & 0.998 \\
 & Average & 0.813 & 0.813 & 0.813 \\
 & Accuracy  & \multicolumn{3}{c}{0.815} \\
\hline
\multirow{7}{*}{DECIGO+ET}
 & PN  & 1.000 & 1.000 & 1.000 \\
 & UL  & 1.000 & 1.000 & 1.000 \\
 & SIS & 0.906 & 0.951 & 0.928 \\
 & CIS & 0.948 & 0.901 & 0.924 \\
 & NFW & 1.000 & 1.000 & 1.000 \\
 & Average & 0.971 & 0.970 & 0.970 \\
 & Accuracy  & \multicolumn{3}{c}{0.971} \\
\hline
\end{tabular}
\end{table}

For single detector configurations, ET achieves an overall Accuracy of $0.677$ and Average $F_1$ of $0.651$, while DECIGO reaches an Accuracy of $0.815$ and Average $F_1$ of $0.813$. As seen in~\ref{tab:overall_metrics_singlecol}, both detectors identify pure noise PN and unlensed UL samples with near perfect precision, and the performance gap mainly arises from discrimination among the three lensed classes. For ET, the Recall of SIS is only $0.172$ with $F_1 = 0.254$, indicating strong confusion when lens dependent waveform differences become less distinguishable in the ET band. This behavior can be understood from the frequency dependent amplification factor in Eq.~\eqref{eq:amplification_factor}: the lensing modulation is governed by the dimensionless frequency parameter $w \propto M_{Lz} f$ (Eq.~\eqref{eq:w}). In the ET frequency range, $w \gg 1$ is more easily satisfied for the adopted lens mass, so the lensing response tends to approach the geometric optics regime, where the wave optics interference patterns are suppressed and the observable differences among $F_{\rm SIS}(w,y)$, $F_{\rm CIS}(w,y)$, and $F_{\rm NFW}(w,y)$ become less prominent in noisy time domain data. Consequently, SIS, CIS, and NFW are more likely to be mutually confused once the effective modulation contrast decreases, even though PN and UL remain well separated. In contrast, DECIGO is sensitive to lower frequencies, where $w \lesssim \mathcal{O}(1)$ is more relevant and wave optics features such as diffraction and interference fringes are more pronounced, leading to more robust lens dependent signatures in the waveform and a higher Average $F_1$. In particular, the NFW class exhibits Precision and Recall close to unity, indicating that the low frequency band carries clearer discriminative structures for certain halo profiles. The comparison of Accuracy and Average $F_1$ therefore reflects the frequency regime dependence of lensing signatures, as controlled by $w$ in Eq.~\eqref{eq:w}, rather than only the overall SNR level.

\begin{figure}
    \centering
    \begin{subfigure}[b]{0.4\textwidth}
        \centering
        \includegraphics[width=\textwidth]{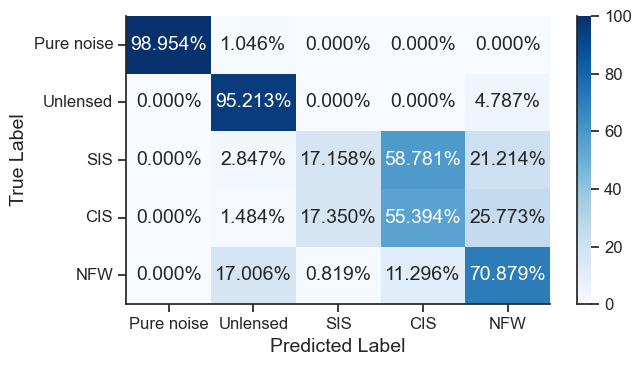}
        \caption{ET}
        \label{subfig:cm_etd}
    \end{subfigure}\hfill
    \begin{subfigure}[b]{0.4\textwidth}
        \centering
        \includegraphics[width=\textwidth]{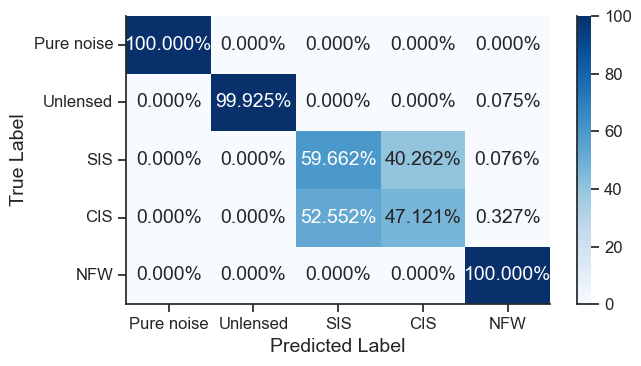}
        \caption{DECIGO}
        \label{subfig:cm_decigo}
    \end{subfigure}\hfill
    \begin{subfigure}[b]{0.4\textwidth}
        \centering
        \includegraphics[width=\textwidth]{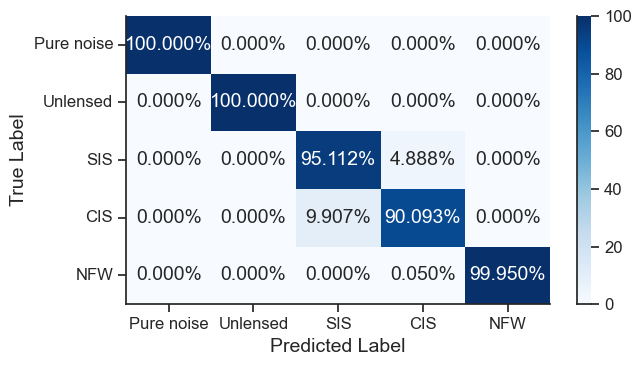}
        \caption{ET+DECIGO}
        \label{subfig:cm_dual}
    \end{subfigure}
\caption{
Confusion matrices for ET, DECIGO, and DECIGO+ET under the full impact parameter range $y \in [0,1]$. The matrices illustrate class-wise prediction accuracy and misclassification patterns among PN, UL, SIS, CIS, and NFW samples.
}
    \label{fig:cm_panel}
\end{figure}

Under the joint configuration DECIGO+ET, the overall Accuracy increases to $0.971$ and Average $F_1$ reaches $0.970$, showing a clear performance gain, as illustrated in~\ref{fig:there_model_metrics} and detailed in~\ref{tab:overall_metrics_singlecol}. The $F_1$ scores of all three lensed classes exceed $0.92$, with SIS and CIS reaching $0.928$ and $0.924$, and NFW nearly perfectly classified. The change in inter-class misclassification can be observed in~\ref{fig:cm_panel}, where cross confusion among lens types is substantially reduced compared to single detector cases. Improvements are also reflected in ROC-AUC and PR-AUC.

This result indicates that wave optics modulation structures in the low frequency band and geometric optics dominated amplitude and phase features in the higher frequency band provide complementary information when combined, enhancing discrimination stability across lens models. Overall, after marginalizing over $y$, the multiband configuration maintains higher classification precision and consistency across the full parameter space.

\subsection{Impact Parameter Dependence}
\label{sec:y_dependence}

\begin{figure}
\centering
\includegraphics[width=0.45\textwidth]{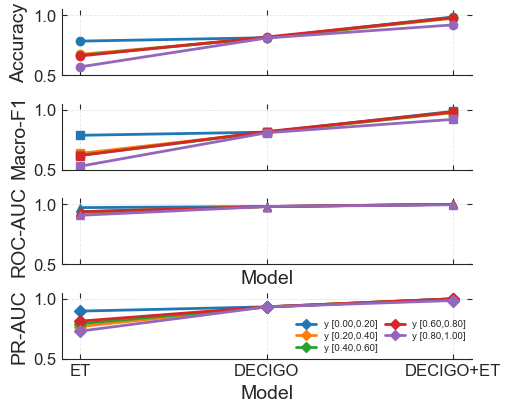}
\caption{
Model performance across five impact parameter bins. Accuracy, Average $F_1$, ROC-AUC, and PR-AUC are shown for ET, DECIGO, and DECIGO+ET, highlighting the dependence of classification performance on the impact parameter $y$.
}
\label{fig:metrics_all_bins}
\end{figure}

To examine the dependence on the impact parameter $y$, we compute Accuracy, Average $F_1$, ROC-AUC, and PR-AUC in five equal-width bins. The metric trends are shown in Fig.~\ref{fig:metrics_all_bins}, and the numerical values are listed in Table~\ref{tab:ybin_metrics_all}.

\begin{table}
\centering
\caption{Model performance in five $y$ bins (each bin: $10^4$ samples per class).}
\label{tab:ybin_metrics_all}
\begin{tabular}{ccccc}
\hline
$y$  & Acc. & Average $F_1$ & ROC-AUC & PR-AUC \\
\hline
\multicolumn{5}{c}{ET} \\
\hline
$[0.0,0.2]$ & 0.786 & 0.788 & 0.973 & 0.897 \\
$[0.2,0.4]$ & 0.675 & 0.637 & 0.925 & 0.768 \\
$[0.4,0.6]$ & 0.668 & 0.621 & 0.938 & 0.791 \\
$[0.6,0.8]$ & 0.662 & 0.619 & 0.938 & 0.813 \\
$[0.8,1.0]$ & 0.571 & 0.530 & 0.908 & 0.729 \\
\hline
\multicolumn{5}{c}{DECIGO} \\
\hline
$[0.0,0.2]$ & 0.814 & 0.814 & 0.981 & 0.932 \\
$[0.2,0.4]$ & 0.812 & 0.812 & 0.981 & 0.932 \\
$[0.4,0.6]$ & 0.814 & 0.813 & 0.981 & 0.931 \\
$[0.6,0.8]$ & 0.820 & 0.819 & 0.982 & 0.933 \\
$[0.8,1.0]$ & 0.812 & 0.809 & 0.981 & 0.932 \\
\hline
\multicolumn{5}{c}{DECIGO+ET} \\
\hline
$[0.0,0.2]$ & 0.988 & 0.988 & 1.000 & 1.000 \\
$[0.2,0.4]$ & 0.977 & 0.977 & 1.000 & 0.999 \\
$[0.4,0.6]$ & 0.981 & 0.981 & 1.000 & 0.999 \\
$[0.6,0.8]$ & 0.983 & 0.983 & 1.000 & 0.999 \\
$[0.8,1.0]$ & 0.922 & 0.921 & 0.996 & 0.985 \\
\hline
\end{tabular}
\end{table}

The ET detector exhibits a clear dependence on $y$. When $y\in[0.0,0.2]$, ET reaches an Accuracy of $0.786$ and Average $F_1$ of $0.788$. As $y$ increases, the performance decreases monotonically, dropping to an Accuracy of $0.571$ and Average $F_1$ of $0.530$ in the $[0.8,1.0]$ bin. The trend is consistent for both Accuracy and Average $F_1$ in~\ref{fig:metrics_all_bins}. The degradation mainly originates from enhanced confusion among the three lensed classes. In the largest bin, the Recall of SIS is only $0.025$ with $F_1 = 0.048$, indicating that its separability becomes very weak at large $y$. In contrast, PN and UL remain well identified across all bins, showing that the performance variation is driven by lens model discrimination rather than signal versus noise separation.

Physically, this trend reflects the dependence of the amplification factor on the impact parameter. As $y$ increases, the source moves farther from the lens center, and the amplification factor $F(w,y)$ approaches unity. In this limit, the lensing modulation becomes progressively weaker and the waveform approaches the unlensed signal. Consequently, the structural differences among the SIS, CIS, and NFW amplification patterns become smaller, reducing the contrast of lens-dependent features in the observed waveform. Since ET operates in a higher-frequency band where the signal is closer to the geometric optics regime, the interference structures are already relatively weak. When $y$ becomes large, the remaining modulation differences among lens models diminish further, which naturally leads to stronger confusion between SIS, CIS, and NFW in the classification task. In contrast, when $y$ is small, the lensing amplification and interference patterns are stronger, producing more distinguishable waveform features and therefore better classification performance.

Compared with ET, DECIGO shows much weaker dependence on $y$. As seen in~\ref{tab:ybin_metrics_all}, its Accuracy remains within $0.812$ to $0.820$, and Average $F_1$ stays around $0.81$. This indicates that modulation structures in the low-frequency band are more robust against variations in $y$. PN and UL are nearly perfectly classified in all bins, and the metrics of SIS and CIS do not show significant degradation with increasing $y$. This behavior can be understood from the physical regime probed by DECIGO. In the lower-frequency band, the dimensionless frequency parameter $w$ in Eq.~\eqref{eq:w} is closer to order unity, placing the signal in the wave-optics regime. In this regime, diffraction and interference structures remain visible even when the impact parameter becomes moderately large, and these patterns depend sensitively on the lens density profile. As a result, the characteristic modulation signatures of different halo models remain detectable over a broad range of $y$, leading to relatively stable classification performance across the five bins.

The joint configuration further enhances stability. For $y\le0.8$, the Accuracy exceeds $0.977$ and Average $F_1$ is close to $0.98$. Even in the $[0.8,1.0]$ bin, the Accuracy remains $0.922$, significantly higher than ET alone. As illustrated in~\ref{fig:metrics_all_bins}, the multiband curves are smoother with reduced fluctuations across bins, indicating complementary information from low and high frequency inputs. Overall, the bin-wise analysis reveals the variation of single-detector performance with lensing strength and demonstrates that the multiband configuration maintains higher consistency and robustness across the full range of $y$.

\section{Conclusion and Outlook}
\label{sec:Conclusion and Outlook}

This study develops a systematic framework for lensed GWs identification under future multiband observations, integrating deep learning with multiband data fusion and performing comprehensive validation on simulated ET and DECIGO datasets. The dataset marginalizes the impact parameter over $y\in[0,1]$, and the performance variation across five $y$ bins is further analysed to reflect classification capability under different lensing strengths. The results show that single detector performance is strongly related to the observational frequency band. ET exhibits stronger discrimination in the small $y$ region, while its performance decreases as $y$ increases and the lensing modulation becomes weaker, leading to enhanced confusion among the three lensed classes. In contrast, DECIGO demonstrates higher stability in the low frequency band, with performance remaining nearly constant across different $y$ bins, indicating stronger robustness against variations in the impact parameter. Under the joint configuration DECIGO+ET, both Accuracy and Average $F_1$ are significantly improved and remain stable across all five $y$ bins, showing clear complementarity between low frequency modulation structures and high frequency amplitude features. The confusion matrices together with ROC-AUC and PR-AUC further confirm that multiband input effectively reduces inter class misclassification. Overall, the results demonstrate that under full marginalization over $y$, multiband observations substantially enhance the stability and reliability of lensed GWs classification and provide a reproducible technical baseline for automated analysis in future observations.

Future work can extend and refine this framework in several directions. On the data side, incorporating more complex lens potentials such as ellipsoidal models \citep{hawken2009gravitational} and compound lens systems \citep{collett2016compound} may improve generalization in realistic lensing environments. On the algorithmic side, future studies may explore architectures combining time domain and frequency domain representations, or introduce attention mechanisms to better capture long timescale modulation features \citep{chen2023long}. The framework can also be extended to joint learning tasks that simultaneously perform lens classification and parameter inference. In addition, evaluation under more realistic observational conditions, including overlapping sources, non stationary noise, and anisotropic detector responses, will be essential to test robustness and scalability. As third generation ground based detectors ET and space detectors DECIGO, LISA \citep{amaro2017laser}, Taiji \citep{ruan2020taiji}, and TianQin \citep{luo2016tianqin} continue to advance, multiband observations are expected to play an increasingly important role in gravitational wave astronomy. In this context, the present framework provides a foundation for rapid lensed GWs identification, lens parameter inference, and statistical studies related to dark matter and cosmology, contributing to the development of precision gravitational wave astrophysics.

\section*{Acknowledgements}

This work was supported by the National Key Research and Development Program of China (Grant No. 2021YFC2203004), the Fundamental Research Funds for the Central Universities Project (Grant No. 2024IAIS-ZD009), the National Natural Science Foundation of China (Grant Nos. 12575072 and 12547101), and the Natural Science Foundation of Chongqing (Grant No. CSTB2023NSCQ-MSX0103). This work made use of \texttt{TensorFlow}~\citep{Abadi_TensorFlow_Large-scale_machine_2015}, \texttt{PyCBC}~\citep{alex_nitz_2024_10473621}, \texttt{GLoW\_public}~\citep{GLoW_public} etc.

\bibliography{wenxian}

\end{document}